\documentclass{elsart41}
\usepackage{graphics}
\usepackage{graphicx}
\usepackage{amssymb}

\begin{document}

\begin{frontmatter}



\title{Na$_\mathrm{x}$CoO$_2$ : Enhanced low-energy excitations of electrons on a 2D triangular lattice}
%

\author[ETH]{M.~Br\"uhwiler\corauthref{cor1}}
\ead{markus.bruehwiler@phys.ethz.ch}
\author[ETH]{B.~Batlogg}
\author[ETH]{S.M.~Kazakov}
\author[PSI]{Ch.~Niedermayer}
\author[ETH]{J.~Karpinski}

\address[ETH]{Laboratory for Solid State Physics, ETH Z\"urich, CH-8093 Z\"urich, Switzerland}
\address[PSI]{Laboratory for Neutron Scattering, Paul Scherrer Institute, CH-5232 Villigen PSI, Switzerland}

\corauth[cor1]{Corresponding author. Tel: +41 44 633 23 39 fax: +41
44 633 10 72}

\newcommand{\fu}{$\mathrm{Na_xCoO_2}$}
\newcommand*{\unit}[1]{\,\mathrm{#1}}
\newcommand{\ce}{\Delta U}  
\newcommand{\CE}{\widetilde{\ce}}  
\newcommand{\sfc}{\gamma}  
\newcommand{\SFC}{\widetilde{\sfc}}  
\newcommand{\Tc}{T_\mathrm{c}}  
\newcommand{\Cp}{C_\mathrm{p}}  
\newcommand{\Hc}{H_\mathrm{c}}  
\newcommand{\svf}{superconducting volume fraction}
\newcommand{\smf}{superconducting mass fraction}

\begin{abstract}

To elucidate the low-energy excitation spectrum of correlated
electrons on a 2D triangular lattice, we have studied the
electrical resistance and specific heat down to $0.5\unit{K}$ and
in magnetic fields up to $14\unit{T}$, in \fu\ samples with a Na
content ranging from $x \approx 0.5$ to $0.82$. Two distinct
regimes are observed: for $x$ from about $0.6$ to $x\approx 0.75$
the specific heat is strongly enhanced, with a pronounced upturn
of $\Cp/T$ below about $10\unit{K}$, reaching $47
\unit{mJ/(mol\,K^2)}$. This enhancement is suppressed in a
magnetic field indicative of strong low-energy spin fluctuations.
At higher Na content the fluctuations are reduced and $\mu$SR data
confirm the SDW ground state below $22\unit{K}$ and the much
reduced heat capacity is field independent.

\end{abstract}

\end{frontmatter}

\newcommand{\fu}{$\mathrm{Na_xCoO_2}$}
\newcommand*{\unit}[1]{\,\mathrm{#1}}
\newcommand{\ce}{\Delta U}  
\newcommand{\CE}{\widetilde{\ce}}  
\newcommand{\sfc}{\gamma}  
\newcommand{\SFC}{\widetilde{\sfc}}  
\newcommand{\Tc}{T_\mathrm{c}}  
\newcommand{\Cp}{C_\mathrm{p}}  
\newcommand{\Hc}{H_\mathrm{c}}  
\newcommand{\svf}{superconducting volume fraction}
\newcommand{\smf}{superconducting mass fraction}


In light of the complexity of the ground state due to geometric
frustration among localized magnetic moments residing on a
triangular lattice,\cite{Ramirez1994} it is of particular interest
to study the low-energy physics of correlated electrons as they
acquire some itineracy on the triangular lattice. \fu\ is
metallic, crystallizes in a layered structure with edge-sharing
CoO$_2$ octahedra, and Co occupies a triangular lattice.
Originally known for its unusually large thermoelectric
effect\cite{Terasaki:12685} this compound exhibits a rich phase
diagram as a function of temperature and Na content. Particularly
notable are the superconductivity in water-intercalated material
($x$ near $0.3$ to $0.4$),\cite{takada:53} a CDW-like
state near $x=0.5$,\cite{foo:247001} and a SDW-like state near
$x=0.8$.\cite{sugiyama:214420,motohashi:064406}

While changing the electron count in the CoO$_2$ layer by
variation of the Na content appears both simple and effective, it
is also to be expected that the Coulomb interaction with the Na
ions will influence the charge dynamics of the itinerant
electrons. This additional complexity becomes even more relevant
because of the partial mobility of the Na ions even at ambient and
modestly elevated temperature, and the resulting possibility for
the Na ions to order. The CDW-like electronic state is a clear
example as it has been associated with a stripe-like ordering of
the Na ions,\cite{zandbergen:024101} and several indications for
Na-ordering at other values of $x$ have been observed
(e.g.~Ref.~\cite{gavilano:100404}). Thus it will be desirable to
identify the influence of the Fermi level shift and that of the
interactions with the Na ions, even as there is reason to expect
some interplay between them.\cite{Indergand2005,Zhang2005} In this
report we focus on the composition range where Na ordering appears
not to dominate the electronic states, at least not on the level
as it does at $x=0.5$.

The specific heat and electrical resistivity were measured in a
Quantum Design PPMS setup with a He-3 insert and a $14\unit{T}$
magnet. Polycrystalline samples were synthesized in a traditional
way and the Na content was confirmed by the unit cell parameters,
in particular by the $c$/$a$ ratio. $\mu^+$SR studies were
performed at the Paul Scherrer Institute and were particularly
helpful to check the magnetic homogeneity of the samples.

\begin{figure}[!ht]
\begin{center}
\includegraphics[width=0.45\textwidth]{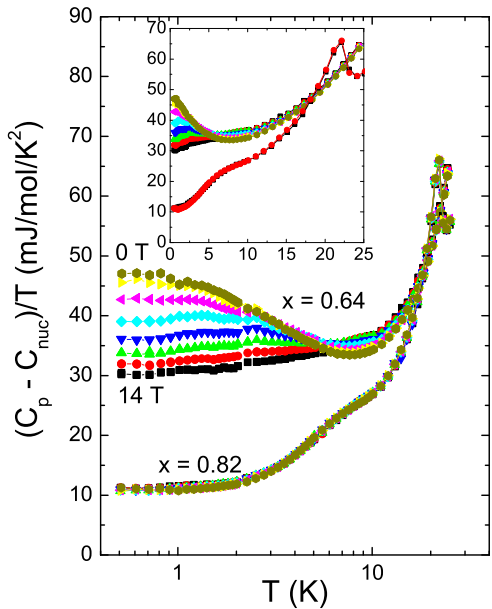}
\end{center}
\caption{\label{fig:CvsTlog} Low-temperature specific heat $\Cp/T$
of two \fu\ samples representing the regions of contrasting
electronic low-$T$ states. The nuclear contribution has been
subtracted for clarity. The $x=0.64$ sample is from the broad
composition region where $\Cp/T$ increases below about
$10\unit{K}$, indicative of strong low energy spin fluctuations,
which are suppressed by a magnetic field. The $x=0.82$ sample
undergoes a SDW-like transition at $22\unit{K}$. The low-$T$ value
of $\Cp/T$ is much lower than in the spin-fluctuation region, and
is not influenced by a magnetic field. The inset shows the same
data on a linear $T$ scale.}
\end{figure}

The essential features of the low-$T$ specific heat for
two representative compositions  are shown in the figure. A well defined nuclear contribution stemming predominantly from Co has been subtracted for clarity. The sample with
$x=0.64$ is representative of samples with a sodium content in the range from $x\approx 0.6$ to
$x\approx0.75$. Zero field  $\mu^+$SR spectra down to $1.7\unit{K}$
in a $x=0.7$ sample indicate the absence of magnetic order. For
this composition range, the specific heat reveals a high density
of low energy excitations, giving rise to an increase of $\Cp/T$
below about $10 \unit{K}$ to reach values as large as $47
\unit{mJ/(mol\,K^2)}$, which has been observed in various studies
(e.g.~\cite{bruehwiler2003,sakurai:2393}). Additional insights
come from measurements in an external magnetic field. In a
magnetic field, these excitations are significantly suppressed,
shown in the figure as the set of data for $0\unit{T}$ to
$14\unit{T}$ with $2\unit{T}$ intervals. The suppression tends to
saturate in high fields and from an extrapolation a total
reduction by about $20 \unit{mJ/(mol\,K^2)}$, corresponding to as much as $43\unit{\%}$, can be estimated.
Apparently, the low-energy excitations are of magnetic origin. The
residual value of $\Cp/T$ is $26$ to $28 \unit{mJ/(mol\,K^2)}$ and
it is still enhanced by about a factor of two compared to the
measured electronic $\Cp/T$ for the superconducting compound at
lower $x$,\cite{jin:217001} or the $x=0.82$ sample also shown in
the figure. For comparison, an early LDA calculation for
Na$_{0.5}$CoO$_2$ gave a $\Cp/T$ of roughly $11
\unit{mJ/(mol\,K^2)}$.\cite{singh:13397} The measured $\Cp/T$ thus
suggests two contributions to the low-energy excitation spectrum.
One with a typical energy scale of about $1 \unit{meV}$, which is
suppressed in a magnetic field, and an other that extends to a
significantly higher energy.

The situation is different at higher Na composition
where a SDW develops, marked by a distinct anomaly at
$22\unit{K}$.\cite{motohashi:064406,sakurai:2393} In the present
study we have focused on the low-$T$ region and the dependence on
a magnetic field. The $\mu^+$SR results of such a sample reveal
the excellent quality in terms of the magnetic volume fraction. A fit of the $\mu^+$SR data using three exponential relaxation functions and a Kubo Toyabe function indicates that over $80\unit{\%}$ of the muons experience a static magnetic field. Below
about $2\unit{K}$ $\Cp/T$ saturates and at $0.5\unit{K}$ amounts
to $11 \unit{mJ/(mol\,K^2)}$. Interestingly, within the
experimental uncertainty of a few percent, this value remains
constant in fields up to $14 \unit{T}$. The same also holds true for the entire temperature range covered: It is  worth noting that neither the transition
temperature at $22\unit{K}$ nor the broad hump near $5$ to
$10\unit{K}$, which has been associated with another magnetic
transition,\cite{sakurai:2393} are modified in a field up to
$14\unit{T}$.

The low-temperature electrical resistivity $\rho$ in the spin fluctuation region ($x=0.7$)
significantly deviates from a $T^2$ behavior. At low $T$, it follows a
temperature dependence $\rho = \rho_0 + \tilde{A} T^a$, with $a$ smaller than $2$. The exponent $a$ starts at about $1.3$ in zero field and increases
rapidly with increasing field reaching about $1.8$ at $14\unit{T}$.

This study was partly supported by the Swiss National Science
Foundation.


\end{document}